\documentclass[twocolumn,showpacs,preprintnumbers,amsmath,amssymb]{revtex4}
\usepackage{mathrsfs}
\usepackage{amssymb}
\usepackage{graphicx}
\usepackage{dcolumn}
\usepackage{bm}
\usepackage{textcomp}

\newcommand\oprod[2]{\ensuremath{|#1\rangle\langle#2|}}

\begin{document}
\title{Making the decoy-state measurement-device-independent quantum key distribution practically useful}
\author{ Yi-Heng Zhou$ ^{1,2}$,Zong-Wen Yu$ ^{1,3}$,
and Xiang-Bin Wang$ ^{1,2,4\footnote{Email
Address: xbwang@mail.tsinghua.edu.cn}\footnote{Also a member of Center for Atomic and Molecular Nanosciences at Tsinghua University}}$}

\affiliation{ \centerline{$^{1}$State Key Laboratory of Low
Dimensional Quantum Physics, Department of Physics,} \centerline{Tsinghua University, Beijing 100084,
People¡¯s Republic of China}
\centerline{$^{2}$ Synergetic Innovation Center of Quantum Information and Quantum Physics, University of Science and Technology of China}\centerline{  Hefei, Anhui 230026, China
 }
\centerline{$^{3}$Data Communication Science and Technology Research Institute, Beijing 100191, China}\centerline{$^{4}$ Jinan Institute of Quantum technology, SAICT, Jinan 250101,
People¡¯s Republic of China}}
\begin{abstract}
 The relatively low key rate seems to be the major barrier to its practical use for the decoy state  measurement device independent quantum key distribution (MDIQKD).
  We present a  4-intensity protocol for the decoy-state MDIQKD that hugely raises the key rate, especially in the case the total data size is not large.
 Also, calculation shows  that our method makes it possible for secure private communication with {\em fresh} keys generated from MDIQKD with a delay time of only a few seconds.
\end{abstract}


\maketitle


\section{Introduction}\label{SecIntro}
One of the most important expected advantage for Quantum key distribution (QKD)\cite{BB84,GRTZ02} is to generate fresh secure keys for instant use, so as to achieve a
higher-level security in private communications. This demands a considerably final key generation rate in a time scale of seconds. The existing technologies
can achieve such a goal if the decoy-state BB84 is applied ~\cite{ILM,H03,wang05,LMC05,wang06,LMC06,AYKI,haya,peng,wangyang,rep,njp}. For example, given the system repetition rate of 1 GHz\cite{gini}, one can make a key rate much higher than the standard of GSM(the Global system for mobile communication), 13K bits per second (bps),  at a rather long distance such as 50 km. This method can keep the unconditional security of QKD  with an imperfect single-photon source~\cite{PNS1,PNS}. However, to patch up the security loophole caused by the limited detection efficiency (including channel loss)~\cite{lyderson}, one has to seek other methods
 such as the so called device independent QKD (DI-QKD)~\cite{ind1} and the measurement-device independent QKD (MDI-QKD) which was based on the idea of entanglement swapping~\cite{ind3,ind2}.  By using the decoy-state method, Alice and Bob can use imperfect single-photon sources~\cite{ind2,wangPRA2013} securely in the MDI-QKD. The decoy-state MDI-QKD has the advantage of getting rid of  all detector side-channel attacks with imperfect single-photon sources. The method has been studied extensively both experimentally~\cite{tittel1,tittel2,liuyang,brazil,loex,Shanghai} and theoretically~\cite{wangPRA2013,MaPRA2012,wangArxiv,lopa,Wang3int, curtty,Wang3improve,Wang3g,Wang1312, WangModel,Xu1406,LaX}.

However, the key rate of the decoy-state MDIQKD is rather low, e.g., in the well known Hefei experiment\cite{Shanghai}, it is 0.018 bps over 200 km, with the set up running for 130 hours. The low key rate seems to be the only barrier to its final practical use. On the other hand, prior art results show that, if the statistical fluctuation is taken into consideration, we  need a large data size so as to reach a considerable final key rate\cite{curtty,MaPRA2012,Xu1406,wangArxiv}. In particular, the number of total pulses at each side $N_t$ is assumed to be larger than $10^{12}$.
It seems to be rather challenging a task to reach a considerable key rate with a small data size, such as $N_t\sim 10^{9}-10^{10}$, which can be done in a time scale of 1 second or a few seconds, given the set-up repetition rate of GHz level\cite{note3,gini}. Note that none of the prior art works can generate a final key in a considerable key rate given a small data size such as $N_t\sim 10^{9}-10^{10}$.

Here we present a method that can produce a key rate much higher than any existing theoretical or experimental results and can generate considerable  key rate in a very short time. Calculation shows that our method can be applied for {\em fresh} key generation with decoy-state MDI-QKD, given the GHz-level set-up repetition rate.

In what follows, we shall first review the decoy-state MDI-QKD and our protocol in section II. We then  take the improved analysis in section III.
There we show a very tricky result that the lower bound of the averaged value of yield and the upper bound phase-flip error rate of all single-photon pairs in both bases can be estimated tightly with observed data
in X basis only.
Based on this fact, we show with explicit formulas that instead of taking the worst-case estimations for the yield and phase-flip error rate of single-photon pairs separately,
we can treat them jointly pointing directly to the worst-case result of the final key rate. We then present the numerical simulation results in section IV.
The results there show  huge advantages of this work in the key rates. The article is ended with a concluding remark.
\section{ protocol}
 We use subscript $A$ or $B$ to denote a source at Alice's side or Bob's side. In our protocol, sources $x_A$ and $y_A$ ($x_B$ and $y_B$) only emit pulses in $X$ basis while source $z_A$ ($z_B$) only emits pulses in $Z$ basis.  The protocol needs four different states  $\rho_{o_A}=\oprod{0}{0}$, $\rho_{x_A}$, $\rho_{y_A}$, $\rho_{z_A}$ ($\rho_{o_B}=\oprod{0}{0}$, $\rho_{x_B}$, $\rho_{y_B}$, $\rho_{z_B}$) respectively.

In photon number space, suppose
\begin{eqnarray}
  \rho_{x_A}=\sum_{k}a_{k}\oprod{k}{k}, &\quad& \rho_{x_B}=\sum_{k}b_{k}\oprod{k}{k}, \label{eq:rhoA} \\
  \rho_{y_A}=\sum_{k}{a_{k}^{\prime}}\oprod{k}{k}, &\quad& \rho_{y_B}=\sum_{k}{b_{k}^{\prime}}\oprod{k}{k}, \label{eq:rhoA} \\
  \rho_{z_A}=\sum_{k}{a_{k}^{\prime \prime}}\oprod{k}{k}, &\quad& \rho_{z_B}=\sum_{k}{b_{k}^{\prime \prime}}\oprod{k}{k}, \label{eq:rhoA}
\end{eqnarray}

We call $x_A$, $x_B$ as well as $y_A$, $y_B$ the decoy sources;  $z_A$, $z_B$ the signal sources, and $o_A$, $o_B$ the vacuum sources.

At each time, Alice will randomly choose source $l_A$ with probability $p_{l_A}$ for $l=o,x,y,z$. Similarly, Bob will randomly choose source $r_B$ with probability $p_{r_B}$ for $r=o,x,y,z$. The emitted pulse pairs (one pulse from Alice, one pulse from Bob)  are sent to the un-trusted third party (UTP).
We shall use notation $lr$ to indicate the two-pulse source when Alice use source $l_A$ and Bob use source $r_B$ to general a pulse pair, e.g., source $xy$ is the source that Alice uses source $x_A$ and Bob uses source $y_B$. Also, here in our protocol, the intensity for pulses in $Z$ basis can be different from those of $X$ basis, this makes more freedom in choosing the intensities and hence further raises the key rate.
Those effective events caused by pulse pairs from source $zz$ will be used for key distillation, while the effective events caused by sources in $X$ basis and vacuum sources will be used to estimate the yield and the phase-flip error rate of the single-photon pulse pairs.

\section{Improved analysis for final key rate}\label{sec:Improved}
We need lower bound value for $s_{11}^Z$, the yield of single-photon pulse pairs in $Z$ basis. However, as discussed by \cite{curtty,Xu1406}, since in actual applications, the number of pulse pairs
in $Z$ is much larger than the number of pulse pairs in X basis, we can use the lower bound of the averaged values of the yield of single-photon pairs in {\em all} bases for the quantity in $Z$ basis only. A very tricky point here is that we can tightly lower bound the yield
of {\em all} single photon pairs  using the observed data in $X-$basis only.
\subsection{Theorems for statistical fluctuation}
We define the counting rate (yield) of pulses of a certain set ${\mathcal C}$ as
\begin{equation}\label{def1}
S_{\mathcal C} =\frac{n_{\mathcal C}}{N_{\mathcal C}}
\end{equation}
where $n_{\mathcal C}$ is the number of valid counts due to pulses in $\mathcal{C}$, and $N_{\mathcal C}$ is the number of pulses in set $\mathcal{C}$.
Actually, in MDI-QKD, we always use pulses pairs from Alice and Bob. So, more strictly speaking, ``pulses" above are actually ``pulse pairs".
Given this definition, we have the following theorem:
\\{\bf Theorem 1:} Suppose set $ {\mathcal{M}}=\{\mathcal{M}_1,\mathcal{M}_2,\cdots ,\mathcal{M}_i,\cdots \mathcal{M}_K\}$ with $K\ge 1$, and any subset $\mathcal{M}_i$ contains $N_{\mathcal{M}_i}$ pulse pairs.  Set $\mathcal{L}=\{\mathcal {L}_1,\mathcal{L}_2,\cdots,\mathcal{L}_i,\cdots,\mathcal{L}_K\}$.  {\em Define} quantity $\langle S_{\mathcal {M}}\rangle_\mathcal{L}= \sum_{i=1}^K c_i S_{\mathcal{L}_i}$ with
$c_i=\frac{N_{\mathcal{M}_i}}{N_{\mathcal{M}}}$.
The following inequality holds
with a probability larger than $1-\epsilon$:
\begin{equation}\label{theorem1}
-\Delta_-\le S_{\mathcal{M}} - \langle S_{\mathcal {M}}\rangle_\mathcal{L}
\le \Delta_{+}
\end{equation}
provided that the following conditions hold for any $i$:\\
i) Set $\mathcal {M}_i$ is a random subset of ${ \mathcal{L}_i}$, i.e., all elements in set ${ \mathcal{L}_i}$ have equal probability to be also an element of  set $\mathcal {M}_i$; $\mathcal M_i\cap \mathcal M_j=\phi$, $\mathcal L_i\cap \mathcal L_j=\phi$ for $i\not= j$ where $\phi$ is the empty set;\\
ii) All elements in ${ \mathcal{L}_i}$ are independent and identical.\\
As shown in \cite{curtty,Shanghai}, the values $\Delta_+,\Delta_-$ can be determined explicitly by using Chernoff bound given the failure probability $\epsilon$.
Though contents of theorem above have been studied and applied elsewhere\cite{curtty,Xu1406,njp}, we believe that
the theorem presented here offers a clearer picture for study of statistical fluctuation in the decoy-state MDI-QKD.
In particular, in considering the averaged value of yield and phase-flip error rate of the single-photon pairs, we don't have to limit them to the average over a certain basis, as was so in Ref.\cite{curtty,Xu1406,LaX}. With our theorem 1, we can use the average over pulses in {\em different} bases.
As demonstrated later in this article, this theorem can help us do calculations more efficiently, e.g., our  theorem 2 and theorem 3.
Obviously, this theorem also holds for the error yields, and hence for the phase-flip error of single-photon pairs, as shall be studied latter.

Theorem 1 shall also apply for the conventional BB84 QKD by just regarding elements of any set there as a single pulse. Here in our application for MDI-QKD, ${\mathcal L}$ shall be a set that contains pulse pairs from several real two-pulse sources. We shall regard pulse pairs of a specific two-mode Fock state from $\mathcal L$ as a set $\mathcal L_i$. Therefore in applying Theorem 1 later in this paper, we shall substitute subscript $i$ by double subscript $mn$ in Theorem 1 and $\mathcal L_{mn}$ is a set that contains all pulse pairs of state $|m\rangle\langle m| \otimes |n\rangle\langle n|$ from $\mathcal L$.
Before any further investigation, we list the following simplified mathematical notations first.
\\(1)  $\mathcal{C}^{lr}$:  the set for  all pulse pairs from  source $lr$; (Sometimes we simply use notation $lr$ for $\mathcal C^{lr}$, if this does not cause any confusion.)
\\(2)   $\mathcal{C}_{mn}$:  the set of pulse pairs of state $|m\rangle\langle m|\otimes |n\rangle\langle n|$ from set $\mathcal{C}$. For example, $\mathcal{C}^{lr}_{mn}$ and $\mathcal L_{mn}$ are sets for pulse pairs of state $|m\rangle\langle m|\otimes |n\rangle\langle n|$ from sets $\mathcal{C}^{lr}$ and $\mathcal L$, respectively.
\\(3)  $s_{mn}^{lr}$:  yield of set $\mathcal{C}^{lr}_{mn}$, i.e., $s_{mn}^{lr} = S_{ \mathcal{C}^{lr}_{mn}}$.
\\  $s^\mathcal {L}_{mn}$: yield of set $\mathcal{L}_{mn}$, i.e., $s^\mathcal {L}_{mn}=S_{\mathcal {L}_{mn}}$.
 \\(4) $S_{lr}$: yield of source  $lr$, i.e. $S_{\mathcal{C}^{lr}}$.

 In any real experimental set-up, the total pulses sent by both sides are finite. In order to extract the secure final key, the effect of statistical fluctuations caused by the finite-size key must be considered. In this case, in general
\begin{equation}
s_{mn}^{lr}\not=   {s_{mn}^{l'r'}}
\end{equation}
 if $l'r'\neq lr$. In this case, to obtain the lower bound value for $s_{11}$ (yield of single-photon pairs) and upper bound value of
 $e_{11}^{ph}$ (phase-flip rate of single-photon pairs),
 one can apply theorem 1 to a suitably chosen set of two pulse sources, $\mathcal D$. As an example we choose\cite{note1}
 \begin{equation}\label{c1}
 \mathcal{D} = \{oo,ox,xo,oy,yo,xx,yy\}.
 \end{equation}
  To apply our theorem, we also need to choose set $\mathcal L$. First, we use
  \begin{equation}\label{l1}
  \mathcal L = \mathcal D.
  \end{equation}
 We can write  the density matrix of any two-pulse source $lr\in \mathcal D$  in the following form
  \begin{equation}\label{rholr}
\rho_{lr}= \sum_{m,n} c^{lr}_{mn}|m\rangle \langle m|\otimes |n\rangle \langle n|.
\end{equation}
 Relating this, we now {\em define} $\langle S_{lr}\rangle_{\mathcal{L} }$ as
\begin{equation}\label{cons1}
\langle S_{lr}\rangle_{\mathcal{L}} = \sum_{m,n} c^{lr}_{mn}s^\mathcal{L}_{mn}.
\end{equation}
 According to this equation,  we can list many equations (constraints) and hence  calculate the lower bound of $s_{11}^{\mathcal L}$ either by formula or by linear programming, as shown in details in the Appendix. Therefore we have the following theorem:

 {\bf Theorem 2}: In the non-asymptotic case, using the observed data (number of counts) of pulses in $X$ basis and pulses from vacuum sources, e.g., pulse pairs from set $\mathcal D$, we can lower bound the yield of single-photon pairs in $X$ basis, the yield of all single photon pairs in in both bases, and also the  yield of those single photon pairs in $Z$ basis only.

 In the Appendix we shall show the explicit formula for lower bound the yield of single-photon pairs in $X$ basis by Eq.(\ref{s11Lx}), and the yield of all single-photon pairs in both $X$ basis and $Z$ basis by  Eq.(\ref{s11Lz}). In particular, the lower bound of the yield of the single-photon pairs is a
 functional of $\mathcal H$ and \begin{equation}\mathcal{H} =a_0 \langle S_{ox}\rangle_{\mathcal L}+b_0 \langle S_{xo}\rangle_{\mathcal L}-a_0 b_0 \langle S_{oo}\rangle_{\mathcal L}.\label{imh}\end{equation}

 Given theorem 2, we can actually deduce the lower bound of yield of  single-photon pairs in $Z$ basis through using the observed data in $X$ basis only, even for the non-asymptotic calculation. {\em This makes it possible to treat the yield of single-photon pairs and the phase-flip error of single-photon pairs jointly} as shown below because both of them are dependent on the same quantity $\mathcal{H}$. We shall take  a detailed study on this very important point below.
  \subsection{Joint study for worst-case result of  key rate}
  We denote $\underline {\bf {\mathcal S}}_{11}(\mathcal H)  $ as the lower bound of $\underline s_{11}^{\mathcal L}$ with a given value $\mathcal H$. Note that $\mathcal{H}$ is defined by Eq.(\ref{imh}). The value $\underline {\bf {\mathcal S}}_{11}(\mathcal H)  $ can be calculated rather tightly if we use constraints
  in Eq.(\ref{contr1}). As shown in the Appendix, the value $\underline {\bf {\mathcal S}}_{11}(\mathcal H)  $ calculated by these constraints can also lower bound
  $\underline s_{11}^{\mathcal L'}$, the yield of all single-photon pairs   because there is a same set of constraints for quantities $\langle S_{lr}\rangle_{\mathcal L'}$, as given by Eqs.(\ref{contr1z},\ref{controx}). Here $\mathcal L' = \mathcal D \cup \mathcal C^{zz}_{11}$. Therefore we shall simply use one notation  $\underline {\bf {\mathcal S}}_{11}(\mathcal H)  $ for both the lower bound of
$\underline s_{11}^{\mathcal L}$ and the lower bound of $\underline s_{11}^{\mathcal L'}$, given $\mathcal H$. Actually, using the method  in Ref.\cite{LaX}, the functional $\underline {\bf {\mathcal S}}_{11}(\mathcal H)  $ can be analytically formulated.

Second, we can also formulate the phase-flip error rate of single-photon pairs for all single-photon pairs\cite{curtty,Xu1406,LaX}. We regard this as  a functional of $\mathcal H$.
 As its original definition in the Pauli channel model\cite{lh,gott}, a phase-flip error is a $\sigma_z=\left(\begin{array}{cc}1&0\\0&-1\end{array}\right)$ error that takes a phase shift in $Z$ basis. We adopt this definition for all qubits in both $X-$basis and $Z-$basis. The $\sigma_z$ errors on qubits in $Z$-basis are not physically detectable, but they do exist. The $\sigma_z$ error on qubits in $X$ basis can be detected, it is just the flipping error in the basis. We can deduce the $\sigma_z$ error rate of all single-photon pairs through measuring the $\sigma_z$ error on those single-photon pairs in $X$ basis only.

Denote the $\sigma_z$-error yield of any set $\mathcal C$ by  $T_{\mathcal{C}}$,  we have $T_{\mathcal{C}}=\frac{\tilde{ n}_{\mathcal C}}{N_{\mathcal C}}$, and
$\tilde{ n}_{\mathcal C}$ is the number or error bits due to set $\mathcal C$.
Denoting $T_{xx}= T_{\mathcal {C}^{xx}}$ we have
\begin{equation}
T_{xx} = \sum_{m,n} c_{mn}^{xx} T^{xx}_{mn}
\end{equation}
where $T^{xx}_{mn}$ is error yield of set $\mathcal C_{mn}^{xx}$.
Taking the same definition of set $\mathcal L'$ as used earlier, we {\em define} quantity $\langle T_{xx} \rangle_{\mathcal L'}$ as
\begin{equation}\langle T_{xx} \rangle_{\mathcal L'} = \sum_{m,n} c_{mn}^{xx} T^{\mathcal L'}_{mn}.
\end{equation}
We can now apply our { theorem 1} to make a non-trivial treatment for error yield and bound the phase-flip error {\em all} single-photon pairs in both bases.
\begin{equation}\label{e11U}
  e^{\mathcal{L'}}_{11}\leq\overline e_{11}^{ph}=\frac{\xi}{a_1 b_1 \underline{\mathcal S}_{11}},
\end{equation}
 and
 $
 \xi= \langle T_{xx}\rangle_{\mathcal L'} - a_0 \langle T_{ox}\rangle_{\mathcal L'}-
 b_0  \langle T_{xo}\rangle_{\mathcal L'}
 + a_0 b_0   \langle T_{oo}\rangle_{\mathcal L'}
 $.
 This gives the phase flip error for all single-photon pairs by using observed data of source $xx$ only.
 To a good approximation, this is also the phase-flip error rate of single-photon pairs in $Z$-basis only, because almost all single-photon pairs are from source $zz$.

 {\bf Theorem 3} Applying our theorem 1, together with Eq.(\ref{e11U}) one can obtain the upper bound of the phase-flip error rate ($\sigma_z$-error rate) for all single-photon pairs by using observed data of source $xx$ only.

 Given the obvious fact that the bit flip error rate must be $5 0\% $ if the bit is caused by source state of $|0\rangle\langle 0|\otimes \rho$ or $\rho\otimes |0\rangle\langle 0|$, we have
 \begin{equation}
 \xi_{xx}= \langle T_{xx}\rangle_{\mathcal L'}-\frac{1}{2} {\mathcal H}.
 \end{equation}
 We can therefore regard the upper bound of $\bar e_{11}^{ph}$ as functional of $\mathcal H$ as
 \begin{equation}
   \bar {\mathcal E}_{11} ({\mathcal H}) = \frac{T_{xx}+\gamma\sqrt{\frac{T_{xx}}{N_{xx}}}-\mathcal H/2}{a_1b_1{\bf \mathcal S}_{11}}\ge \bar e_{11}^{ph}.
 \end{equation}

Therefore,  we have the following key rate formula as a functional of $\mathcal H$
\begin{equation}\label{KeyRateRZX}
\begin{aligned}
  &\mathcal R(\mathcal H) =p_{z_A}p_{z_B}\\
   &\cdot\{a_1^{\prime \prime Z} b_1^{\prime \prime Z} \underline {\bf \mathcal S}_{11}(\mathcal  H)[1-H(\bar {\mathcal{E}}_{11}(\mathcal H)]-f S_{zz}H(E_{zz})\},
\end{aligned}
\end{equation}
where $f$ is the error correction inefficiency and $E_{zz}$ is the observed bit error rate for source $zz$.
The final key is simply the worst-case result of $\mathcal R (\mathcal H) $ over all possible values for $\mathcal H$, i.e.
\begin{equation}\label{KeyRateRZX1}
  R= \min_{\mathcal H\in \mathcal I} \mathcal R({\mathcal H}).
\end{equation}
According to Eq.(\ref{controx}), we have
$\mathcal I =[h-\delta,h+\delta]$ and
\begin{equation}\label{hrange}
\begin{aligned}
&h=a_0S_{ox}+a_0S_{xo}-a_0^2S_{oo}\\
& \delta=a_0\gamma \sqrt{\frac{S_{ox}+S_{xo}}{N_{ox}}}-a_0^2\gamma\sqrt{\frac{S_{oo}}{N_{oo}}}
\end{aligned}
\end{equation}
given the symmetric set-up that satisfies $a_0=b_0$, $N_{xo}=N_{ox}$. We shall use such a symmetric case in our numerical simulation.
\section{Numerical simulation}\label{sec:SimulationNew}
In this section, we present some numerical simulations in comparison with the best known prior art results Ref.\cite{Xu1406,LaX}. We focus on the symmetric case where the two channel transmissions from Alice to UTP and from Bob to UTP are equal. We also assume that the UTP's detectors are identical, i.e., they have the same dark count rates and detection efficiencies, and their detection efficiencies do not depend on the incoming signals. Also, we assume
\begin{equation}
a_k=b_k,\; a_k'=b_k',\; a_k''=b_k''
\end{equation}
for all $k$. And $p_{l_A}=p_{r_B}$ for any $l=r$.

We shall estimate what values would be probably observed for the yields and error yields in the normal cases by the linear models as in~\cite{wang05,ind2,MaPRA2012,WangModel,Xu1406}. We shall assume 2 types of detectors. Experimental conditions and parameters of detectors~\cite{tittel1,UrsinNP2007}  are listed in Table~I. For the second type of detector, we assume $40\%$ detection and $10^{-7}$ dark count rate, in Fig.(\ref{b9},\ref{b11},\ref{c13}).  In Fig.(\ref{c13}), we assumed the alignment error probability to be $e_d=1\%$. With these, the yields and error rate can be simulated~\cite{MaPRA2012,WangModel}. We assume a coherent state for all sources. The density matrix of the coherent state with intensity $\mu$ can be written into $\rho=\sum_{k}\frac{e^{-\mu}\mu^k}{k!} \oprod{k}{k}$. We calculate the key rate using Eq.(\ref{KeyRateRZX},\ref{KeyRateRZX1})

\subsection{Numerical results}
Here we first take a simple treatment using normal distribution  in order to make a fair comparison with the prior art results\cite{Xu1406,LaX}.
This means that we can set $\Delta_+=\Delta_- = \gamma \sqrt {\frac {S_{\mathcal{C}}}{N_{ \mathcal {C}}}}$, and $\gamma=5.3$ given the failure probability $\epsilon = 10^{-7}$. We emphasize that no matter we use Chernoff bound or the simple treatment, the conclusion that our method can hugely improve the key rates does not change. We use the full parameter optimizations for all protocols~\cite{Xu1406,LaX}.
\begin{figure}
  \includegraphics[width=240pt]{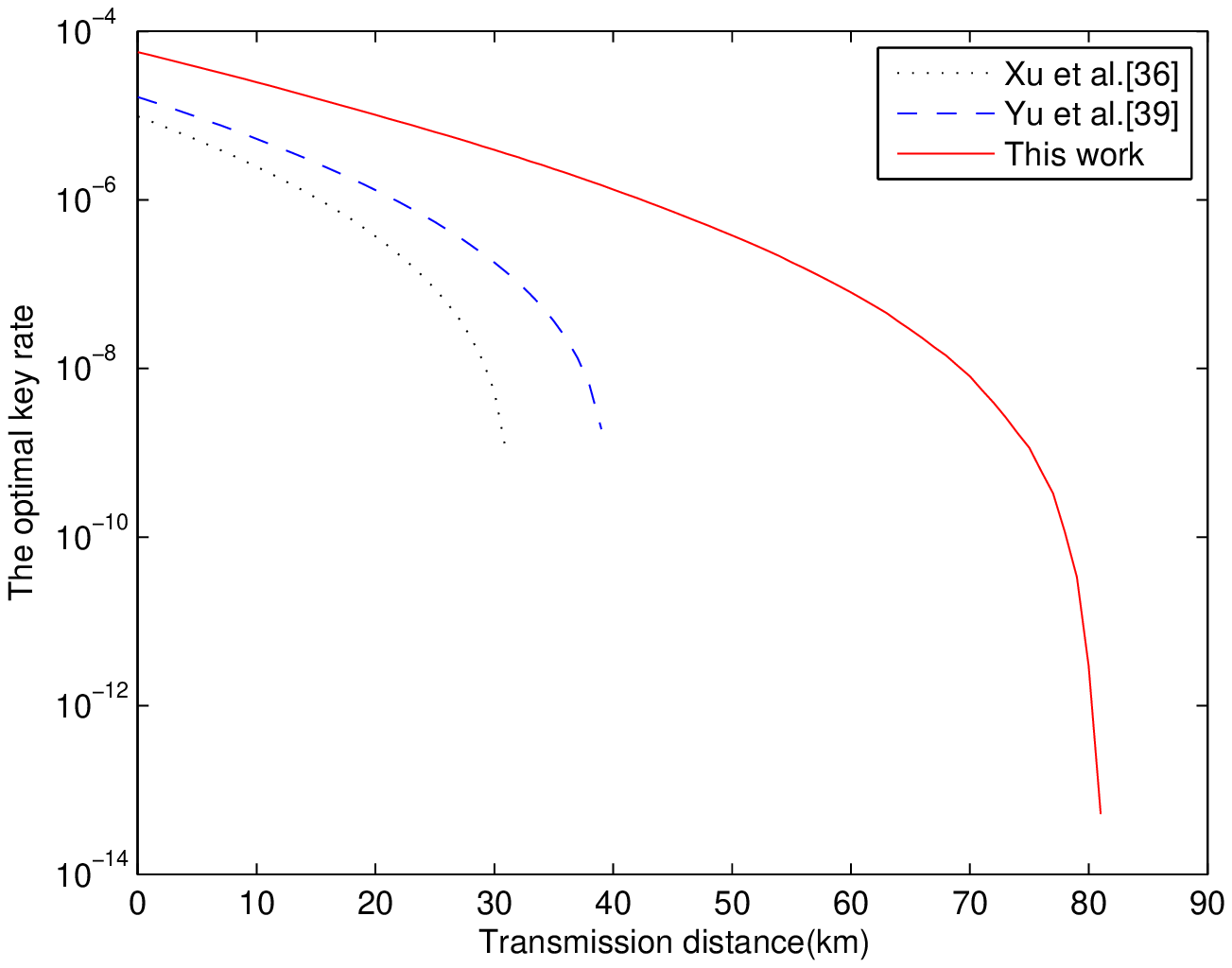}\\
  \caption{(Color online) The optimized key rates (per pulse pair) versus transmission distance by different methods  with device parameters being given by line $a$ of Table I\ref{tab:list}. Here we set the total number of pulses at each side $N_t=10^{10}$ and the failure probability $\epsilon=10^{-7}$.}\label{a10}
\end{figure}

\begin{figure}
  \includegraphics[width=240pt]{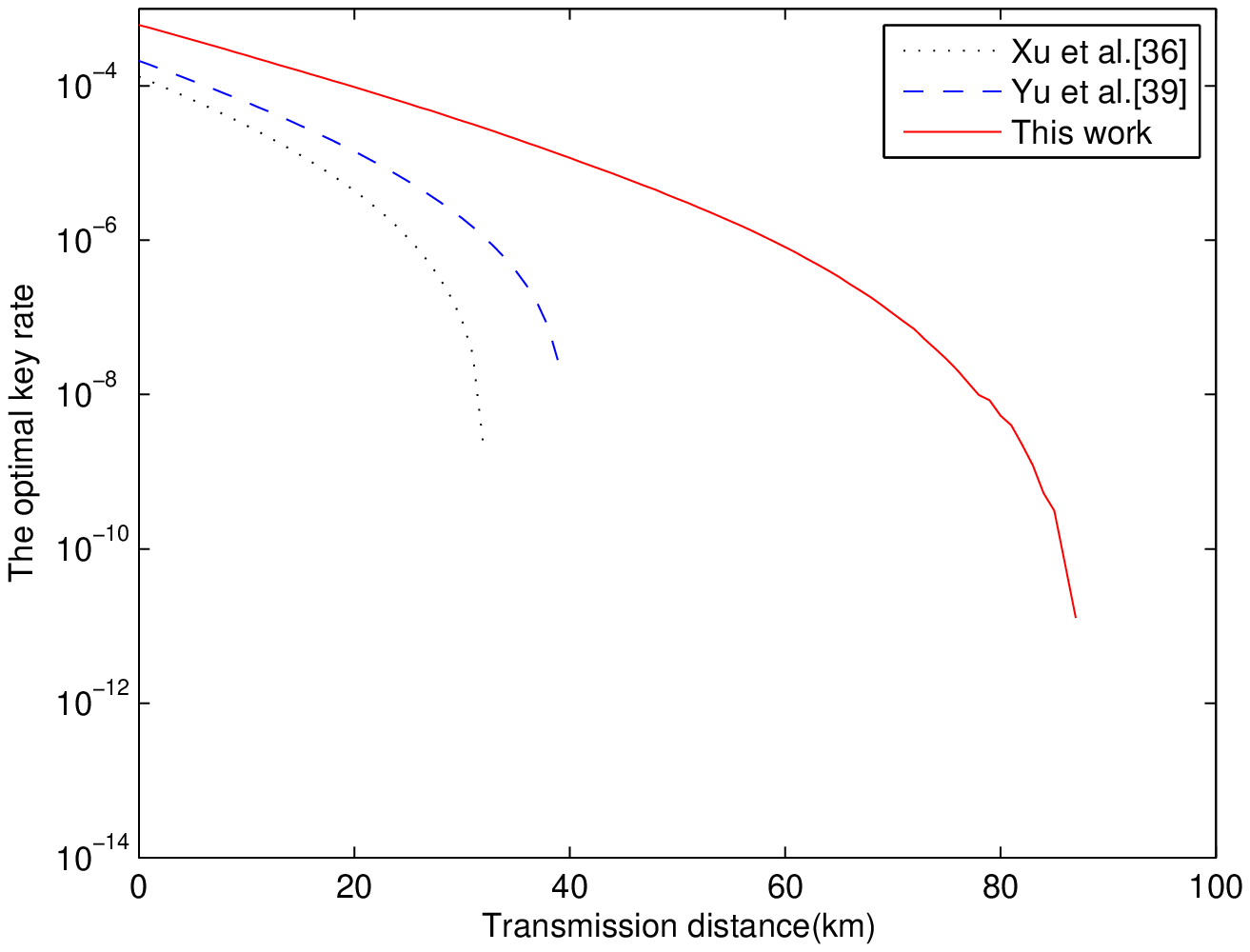}\\
  \caption{(Color online) The optimized key rates (per pulse pair) versus transmission distance by different methods
  with device parameters being given by line $b$ of Table \ref{tab:list}. Here we set the total number of pulses at each side $N_t=10^{9}$ and the failure probability $\epsilon=10^{-7}$.}\label{b9}
\end{figure}

\begin{figure}
  \includegraphics[width=240pt]{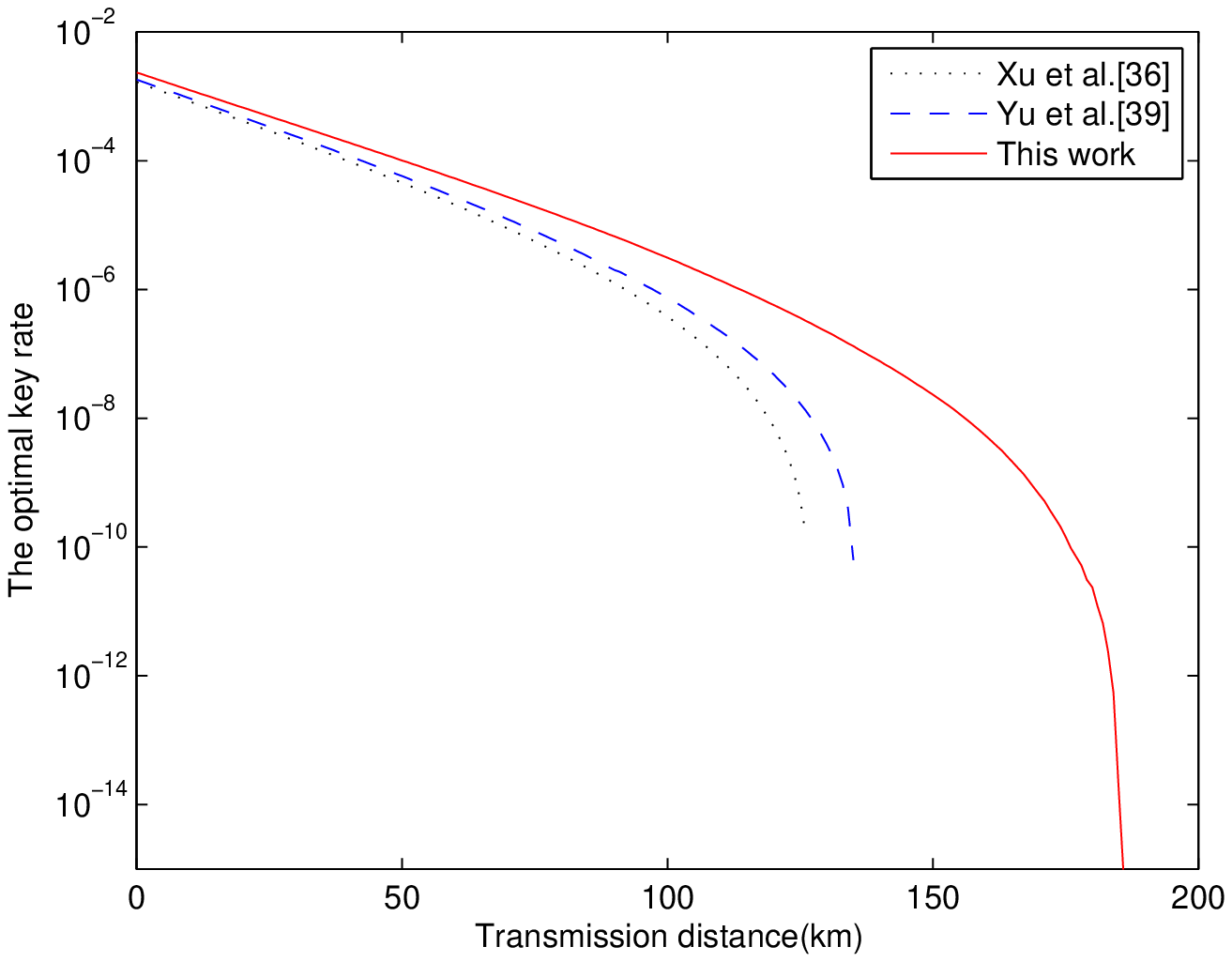}\\
  \caption{(Color online) The optimized key rates (per pulse pair) versus transmission distance by different methods with device parameters being given by line $b$ of Table \ref{tab:list}. Here we set the total number of pulses at each side $N_t=10^{11}$ and the failure probability $\epsilon=10^{-7}$.}\label{b11}
\end{figure}

\begin{figure}
  \includegraphics[width=240pt]{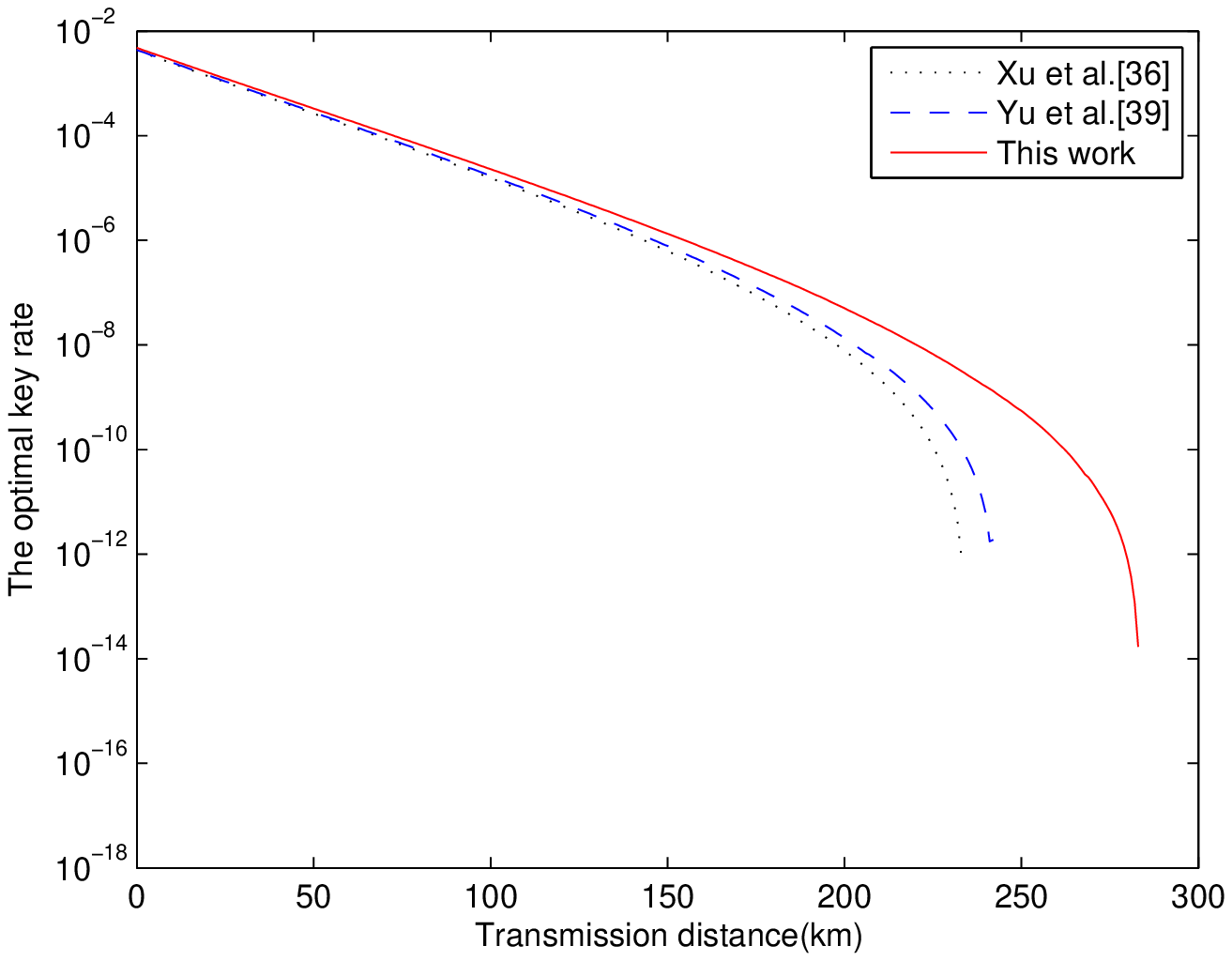}\\
  \caption{(Color online) The optimized key rates (per pulse pair) versus transmission distance by different methods with device parameters being given by line $c$ of Table \ref{tab:list}. Here we set the total number of pulses at each side $N_t=10^{13}$ and the failure probability $\epsilon=10^{-7}$.}\label{c13}
\end{figure}

\begin{table}
\begin{ruledtabular}
\begin{tabular}{cccccc}
   & $e_0$ & $e_d$ & $p_d$ & $\eta_d$ & $f_e$   \\
  \hline
  $a$ &0.5 & 1.5\% & $6.02\times 10^{-6}$ & $14.5\%$ & $1.16$  \\
  $b$ &0.5 & 1.5\% & $10^{-7}$ & $40\%$ & $1.16$  \\
  $c$ &0.5 & 1\% & $10^{-7}$ & $40\%$ & $1.16$  \\
\end{tabular}
\caption{\label{tab:list}Device parameters used in numerical simulations. $p_d$: the dark count rate. $\eta_d$: the detection efficiency of all detectors. $f_e$: the error correction inefficiency. We shall use the device parameters of line $a$ for the calculation of Fig. \ref{a10} and Table \ref{tab:gash}; line $b$ for calculation of Fig. \ref{b9}, Fig. \ref{b11} Fig. \ref{50kmyu}, Fig. \ref{50kmzhou}, and Table \ref{tab:supper}, and line $c$ for the calculation of Fig. 4.}
\end{ruledtabular}
\end{table}
 Fig.(\ref{a10},\ref{b9},\ref{b11},\ref{c13}) make a clear view that our method improves the final key rate and transmission distance drastically,
and the advantage is much more outstanding when the data size is smaller.

In these figures, the black dotted curve is the key rate obtained from the method in Ref.(\cite{Xu1406}), where fluctuations of each sources are treated
separately.  The blue dashed curve is the improved key rate using the method of our previous work in Ref.\cite{LaX}, and the red solid curve is
the result of  this work.

Table II and Table III list the key rates at different distances for different protocols. From the tables we can see that our method at this work can improve the key rate by 20-100 times in a typical parameter set. Table IV lists the optimized parameters of our method at 40 km for our result in Table I and 50 km for our result in Table III.

\begin{table}
\begin{ruledtabular}
\begin{tabular}{c|cccc}
  \rm{Distance} & 30 km & 35 km & 40 km & 50 km \\
  \hline
  Xu et al.\cite{Xu1406} & $5.32\times 10^{-9}$ & - & - & - \\
  Yu et al.\cite{LaX} & $1.81\times 10^{-7}$ & $3.66\times 10^{-8}$ & - & -\\
  \rm{This work} & $3.93\times 10^{-6}$ & $2.33\times 10^{-6}$ & $1.33\times 10^{-6}$ & $3.78\times 10^{-7}$ \\
\end{tabular}
\caption{\label{tab:gash} Comparison of key rates at different distance (standard fiber)  for calculations done in  Fig.(\ref{a10}).}
\end{ruledtabular}
\end{table}

\begin{table}
\begin{ruledtabular}
\begin{tabular}{c|ccc}
  \rm{Distance} & 35 km & 50 km & 60 km\\
  \hline
  Xu et al.\cite{Xu1406} & - & - & - \\
  Yu et al.\cite{LaX} & $4.02\times 10^{-7}$ & - & - \\
  \rm{This work} & $2.07\times 10^{-5}$ & $3.44\times 10^{-6}$ & $8.16\times 10^{-7}$ \\
\end{tabular}
\caption{\label{tab:supper} Comparison of key rate at different distance (standard fiber) for calculations done in Fig.(\ref{b9}).}
\end{ruledtabular}
\end{table}

\begin{table}
\begin{ruledtabular}
\begin{tabular}{ccccccc}
  &$\mu_x$ & $\mu_y$ & $\mu_z$ & $p_x$ & $p_y$ & $p_z$ \\
  \hline
  \ref{tab:gash} & 0.071 & 0.212 & 0.280 & 0.357 & 0.121 & 0.479 \\
  \ref{tab:supper} & 0.078 & 0.241 & 0.252 & 0.398 & 0.138 & 0.423 \\
\end{tabular}
\caption{\label{tab:opptm}List of optimized parameters used in numerical simulations by the method of \rm{this work}.  Line \ref{tab:gash}: for 40 km in Table \ref{tab:gash}; Line \ref{tab:supper}: for 50 km in Table \ref{tab:supper}. }
\end{ruledtabular}
\end{table}
\subsection{Results with higher security}
We have also calculated the key rates by using Chernoff bound\cite{curtty}, the conclusion that our method here can hugely improve the efficiency of
MDIQKD keeps unchanged.
Consider the Hefei experiment. For a fair comparison, we strictly use Chernoff bound with failure probability $10^{-10}$. And also in the final key rate calculation, we use the yield lower bound values of those single-photon pairs in $Z$ basis only.
Say, we shall replace  ${\bf \mathcal S}_{11}(\mathcal  H)$ and $\bar {\mathcal{E}}_{11}(\mathcal H)$ in Eq.(\ref{KeyRateRZX}) by
${\kappa_s\bf \mathcal S}_{11}(\mathcal  H)$ and $\kappa_e(\bar {\mathcal{E}}_{11}(\mathcal H)$, respectively, where the factors $\kappa_s,\;\kappa_e$  are in ranges around  1 due to the possible fluctuations between
all single-photon pulses and those from $zz$ source, for the quantities yield and phase-flip error rate, respectively. We use results of Ref.\cite{Shanghai} for bounds of $\kappa_s,\;\kappa_e$.

  Also, according to the observed data there\cite{Shanghai}, we used a linear loss model to estimate the actual over all loss in the experiment. Assuming the same device parameter, we make the optimization by using our 4-intensity protocol. We obtain a final key rate of 0.98 bit per second (bps), which is more than 50 times higher than the reported experimental result, 0.0177 bps.

\begin{figure}
  \includegraphics[width=240pt]{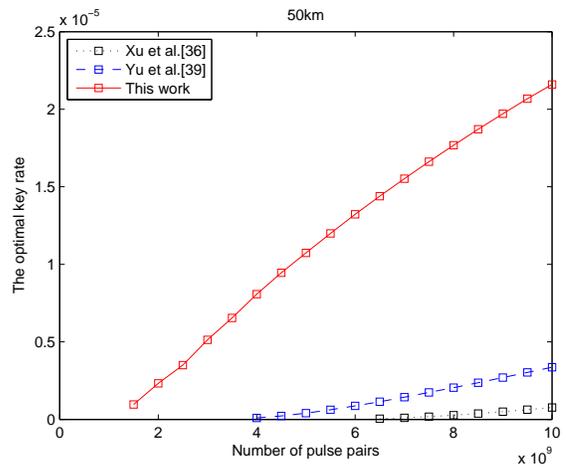}\\
  \caption{(Color online) The optimized key rates (per pulse pair) versus different total number of pulse pairs $N_t$ for each protocols at the distance of 50 km. Here the failure probability is $\epsilon = 10^{-10}$, with the device parameters being listed in line $b$ of Table \ref{tab:list}. We strictly use Chernoff bound in the calculation.}\label{50kmyu}
\end{figure}

\begin{figure}
  \includegraphics[width=240pt]{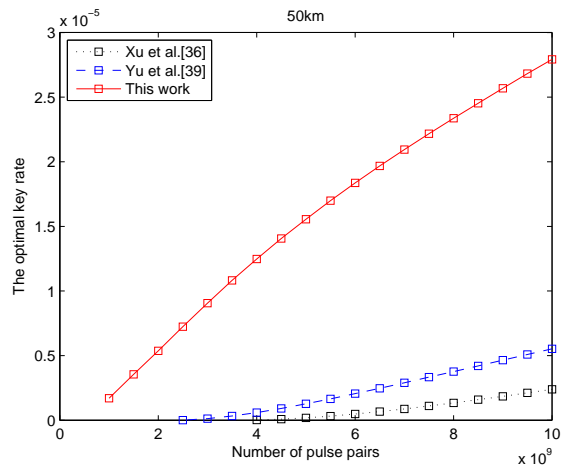}\\
  \caption{(Color online) The optimized key rates (per pulse pair) versus different total number of pulse pairs $N_t$ for each protocols at the distance of 50 km calculated by the Normal distribution approximation. Here the failure probability is also set to be $10^{-10}$, with the device parameters being listed in line $b$ of Table \ref{tab:list}.}\label{50kmzhou}
\end{figure}

Furthermore, consider the possibility fresh key application by our method. The standard of GSM  requests a transmission rate in 13 kbps. Taking this standard, we calculate   the key rates of different protocols with various total number of pulses from $10^{9}$ to $10^{10}$, at a fixed distance of 50 km, as shown in Fig.(\ref{50kmyu}). From this figure we can see that our protocol can fulfill the task of private mobile phone communication with only less than 5.9 seconds delay if the system repetition rate is 1 GHz \cite{expv}. This is an impossible job for all prior art protocols. For a comparison, we also did the same calculation by Normal distribution approximation as used earlier, with the failure probability $10^{-10}$, in Fig.(\ref{50kmzhou}). Here, we find the delay time of our protocol is only about 4.1 seconds if the system repetition rate is 1 GHz.

\section{Concluding remark}\label{sec:Conclusion}
In real set-ups of MDI-QKD, the effects of statistical fluctuations caused by the finite-size key must be considered. In the statistical analysis, earlier works\cite{curtty,Xu1406,LaX} used the simple worst-case calculation for single-photon yield and the phase-flip error rate separately, leaving the problem of difference between the error rate in $X$ basis and the phase-flip error in $Z$ basis. Here we used the more economic worse-case estimation pointing directly to the final key rate, and calculate the yield and phase-flip error rate directly $Z$-basis using the data in $X$-basis only. These improved the key rate drastically. Also, here in our protocol intensities of pulses at different bases can be different, this further improves the key rate. Also, we have shown that actually both the yield and the phase-flip error rate of single-photon pairs can be calculated directly for {\em all} single-photon pairs using observed data in $X$ basis only.   As shown in  the numerical simulations, the results obtained with our improved methods are much better than the results obtained before. In short, we have proposed a method that is much more efficient than all
 known methods for improving key rate in the decoy-state MDI-QKD. Our method has actually made the decoy-state MDI-QKD immediately useful in practice.

In our calculation, we have chosen a special of sources for $\mathcal D$ in Eq.(7). As was pointed out already, there are other choices, e.g., $\{oo,ox,xo,oy,yo,xx,xy,yx\}$\cite{LaX}, linear programming and so on. The method here can also be applied to the traditional decoy-state BB84 protocol, say Alice has vacuum source, source $x,y$ in $X$ basis and source $z$ in $Z$ basis as the signal source. They use vacuum source and sources $x,y$ to calculate the single photon yield and phase-flip error rate and use source $z$ for key distillation. And also one can treat the single photon yield and phase-flip error rate jointly, taking the optimization directly pointing to the final key rate. This will be reported elsewhere.
\\{\bf Acknowledgement} XBW proposed this work and presented  key-rate analysis. YHZ and ZWY did the numerical test. YHZ and XBW wrote the paper. We
acknowledge the financial support in part by the 10000-Plan of Shandong province (Taishan Scholars),
National High-Tech Program of China grant No. 2011AA010800 and 2011AA010803,
NSFC grant No. 11474182, 11174177 and 60725416, and the key R$\&$D Plan Project of Shandong Province,  grant No. 2015GGX101035.
\\{\em Note added:} Several months after we announced this work in arXiv 1502:01262 (2015), our method proposed in this work has been experimentally implemented very recently\cite{camb}. There, 1 GHz source repetition rate is demonstrated with pulses' coherence length of 0.035 ns.
\section{Appendix}
Given set ${\mathcal L} ={\mathcal D}$, using constraints
\begin{equation}
\langle S_{lr}\rangle_{\mathcal{L}} = \sum_{m,n} c^{lr}_{mn}s^\mathcal{L}_{mn}.
\end{equation}
as shown in Eq.(\ref{cons1}), we have
 \begin{widetext}
\begin{equation}\label{s11Lx}
\begin{aligned}
& s_{11}^{\mathcal L}\ge \underline s_{11}^{\mathcal L}= \frac{[a_{1}^{\prime} b_2^{\prime} \langle S_{xx}\rangle_{\mathcal L} +a_1 b_2 a_0^{\prime} \langle S_{oy}\rangle_{\mathcal L} +a_1 b_2 b_0^{\prime} \langle S_{yo}\rangle_{\mathcal L}] - [a_1 b_2 \langle S_{yy}\rangle_{\mathcal L} + a_1 b_2 a_0^{\prime} b_0^{\prime} \langle S_{oo}\rangle_{\mathcal L}]-a_1^{\prime} b_2^{\prime} \mathcal{H}}{a_1 a_1^{\prime} (b_1 b_2^{\prime}-b_1^{\prime} b_2)},\\
& {\rm and}\;\mathcal{H} =a_0 \langle S_{ox}\rangle_{\mathcal L}+b_0 \langle S_{xo}\rangle_{\mathcal L}-a_0 b_0 \langle S_{oo}\rangle_{\mathcal L}.
\end{aligned}
\end{equation}
\end{widetext}
Since quantities $\langle S_{lr}\rangle_{\mathcal L}$ are not exactly determined, we can only find out the lower bound for for $\underline {s}^{\mathcal{L}}_{11}$ with constraints for fluctuations given by Eq.({\ref{theorem1}}) in our theorem 1.
We can rewrite $\mathcal L$ in $\mathcal L =\{\mathcal L_{mn}|m=0,1,2,\cdots ; n=0,1,2,\cdots\}$  where the subset $\mathcal L_{mn}$ ($\mathcal L'_{mn}$) is for all pulse pairs in state $|m\rangle\langle m|\otimes |n\rangle\langle n|$ from set $\mathcal L$ ($\mathcal M$). We immediately find that
for any source $lr\in \mathcal D$, $\mathcal C^{lr}_{mn}\in \mathcal L_{mn}$ and also $\mathcal C^{lr}_{mn}\in \mathcal L_{mn}$. Regarding each $\mathcal C^{lr}_{mn}$ as $\mathcal M$ and  $\{\mathcal C^{lr}_{mn}\}$ as $\{\mathcal M_i\}$ in our theorem, we find that condition 1 in theorem 1 holds. Moreover, one can easily find that all conditions in our theorem 1 hold  for sets $\{\mathcal{C}^{lr},  \mathcal{L}\}$   above. Therefore
we can use Eq.(\ref{theorem1}) for constraints of fluctuations.
We shall use the following constraints:
\begin{widetext}
\begin{equation}\label{contr1}
\begin{aligned}
& N_{lr} S_{lr} +\gamma \sqrt{ N_{lr} S_{lr}} \ge N_{lr}\langle S_{lr} \rangle_{\mathcal L} \ge N_{lr} S_{lr} -\gamma \sqrt{ N_{lr} S_{lr}}\;;\; {\rm for\; any}\; lr\in \mathcal D\\
&  N_{yo}\langle S_{yo}\rangle_{\mathcal L} +N_{oy}\langle S_{oy}\rangle_{\mathcal L} \ge N_{yo} S_{yo} + N_{oy} S_{oy} - \gamma\sqrt{N_{yo} S_{yo} +N_{oy} S_{oy}}\\
& N_{xx}\langle S_{xx}\rangle_{\mathcal L}+ N_{yo}\langle S_{yo}\rangle_{\mathcal L} +N_{oy}\langle S_{oy}\rangle_{\mathcal L} \ge
 N_{xx} S_{xx}+ N_{yo} S_{yo} +N_{oy} S_{oy}- \gamma \sqrt{N_{xx} S_{xx}+ N_{yo} S_{yo} +N_{oy} S_{oy}}\\
& N_{yy}\langle S_{yy}\rangle_{\mathcal L} + N_{oo}\langle S_{oo}\rangle_{\mathcal L}\le N_{yy} S_{yy} + N_{oo} S_{oo} + \gamma\sqrt{N_{yy} S_{yy} + N_{oo} S_{oo}}\\
\end{aligned}
\end{equation}
and \begin{equation}\label{controx}
 N_{ox} S_{ox}+ N_{xo} S_{xo} + \gamma \sqrt{N_{ox} S_{ox}+ N_{xo} S_{xo}} \ge N_{xo}\langle S_{xo}\rangle_{\mathcal L} +N_{ox}\langle S_{ox}\rangle_{\mathcal L} \ge N_{ox} S_{ox}+ N_{xo} S_{xo} - \gamma \sqrt{N_{ox} S_{ox}+ N_{xo} S_{xo} }
\end{equation}
\end{widetext}
The  first line of Eq.(\ref{contr1}) gives individual ranges of statistical fluctuations related to each separate sources, the other lines are {\em joint} constraints among different sources, as was studied in detail in Ref.(\cite{LaX}).

Second, we use set
  \begin{equation}\label{lp2}
  \mathcal L' = \mathcal D \cup \mathcal C^{zz}_{11}.
  \end{equation}
  and quantities
  \begin{equation}\label{consp2}
\langle S_{lr}\rangle_{\mathcal{L'}} = \sum_{m,n} c^{lr}_{mn}s^\mathcal{L'}_{mn}.
\end{equation}
 Note that $\mathcal L'$ is simply the set for pulse pairs from sources in $\mathcal D$ and single-photon pairs from source $zz$.  As shown in Ref.\cite{wangPRA2013}, the states in polarization space for single-photon pairs in $X$ basis or in $Z$ basis are identical. Both of them are $\frac{1}{4}I$. Similar to the study above for set $\mathcal L$, it is easy to see that all conditions in our theorem 1 hold  for sets  $\{\mathcal{C}^{lr},  \mathcal{L'}\}$ above.
Similar to Eq.(\ref{s11Lx}), we have the lower bound
of $s_{11}^{\mathcal L'}$ by
 \begin{widetext}
\begin{equation}\label{s11Lz}
\begin{aligned}
& s_{11}^{\mathcal L'}\ge \underline s_{11}^{\mathcal L'}
= \frac{[a_{1}^{\prime} b_2^{\prime} \langle S_{xx}\rangle_{\mathcal L'}
 +a_1 b_2 a_0^{\prime} \langle S_{oy}\rangle_{\mathcal L'} +a_1 b_2 b_0^{\prime}
 \langle S_{yo}\rangle_{\mathcal L'}]
  - [a_1 b_2 \langle S_{yy}\rangle_{\mathcal L'} + a_1 b_2 a_0^{\prime} b_0^{\prime} \langle S_{oo}\rangle_{\mathcal L'}]
-a_1^{\prime} b_2^{\prime} \mathcal H'}{a_1 a_1^{\prime} (b_1 b_2^{\prime}-b_1^{\prime} b_2)},\\
& {\rm and}\;\mathcal{H'} =a_0 \langle S_{ox}\rangle_{\mathcal L'}+b_0 \langle S_{xo}\rangle_{\mathcal L'}-a_0 b_0 \langle S_{oo}
\rangle_{\mathcal L'}.
\end{aligned}
\end{equation}
\end{widetext}
And we shall use the following constraints from Eq.(\ref{theorem1}) in theorem 1:
\begin{widetext}
\begin{equation}\label{contr1z}
\begin{aligned}
& N_{lr} S_{lr} +\gamma\sqrt{ N_{lr} S_{lr}} \ge N_{lr}\langle S_{lr} \rangle_{\mathcal L'} \ge N_{lr} S_{lr} -\gamma \sqrt{ N_{lr} S_{lr}}\;;\; {\rm for\; any}\; lr\in \mathcal D\\
&  N_{yo}\langle S_{yo}\rangle_{\mathcal L'} +N_{oy}\langle S_{oy}\rangle_{\mathcal L'} \ge N_{yo} S_{yo} + N_{oy} S_{oy} - \gamma\sqrt{N_{yo} S_{yo} +N_{oy} S_{oy}}\\
& N_{xx}\langle S_{xx}\rangle_{\mathcal L'}+ N_{yo}\langle S_{yo}\rangle_{\mathcal L'} +N_{oy}\langle S_{oy}\rangle_{\mathcal L'} \ge
 N_{xx} S_{xx}+ N_{yo} S_{yo} +N_{oy} S_{oy}- \gamma \sqrt{N_{xx} S_{xx}+ N_{yo} S_{yo} +N_{oy} S_{oy}}\\
& N_{yy}\langle S_{yy}\rangle_{\mathcal L'} + N_{oo}\langle S_{oo}\rangle_{\mathcal L'}\le N_{yy} S_{yy} + N_{oo} S_{oo} + \gamma\sqrt{N_{yy} S_{yy} + N_{oo} S_{oo}}\\
\end{aligned}
\end{equation}
and \begin{equation}\label{controxz}
 N_{ox} S_{ox}+ N_{xo} S_{xo} + \gamma \sqrt{N_{ox} S_{ox}+ N_{xo} S_{xo}} \ge N_{xo}\langle S_{xo}\rangle_{\mathcal L'} +N_{ox}\langle S_{ox}\rangle_{\mathcal L'} \ge N_{ox} S_{ox}+ N_{xo} S_{xo} - \gamma \sqrt{N_{ox} S_{ox}+ N_{xo} S_{xo} }
\end{equation}
\end{widetext}
Therefore we arrive at the conclusion below:

 {\em In the non-asymptotic case, the yield of single-photon pairs in $X$ basis is lower bounded by Eq.(\ref{s11Lx}), and the yield of all single-photon pairs in both $X$ basis and $Z$ basis is lower bounded  Eq.(\ref{s11Lz}).}

To a good approximation, we can regard the lower bound of $\underline {s}^{\mathcal{L'}}_{11}$ above as the lower bound of the yield of single-photon pairs in $Z$ basis, since in our protocol most of (actually, almost all) single-photon pairs which can cause effective events are produced in $Z$ basis.

Also, compare Eq.(\ref{s11Lx}) and Eq.(\ref{s11Lz}) we find that, the right side of Eq.(\ref{s11Lz}) actually has the same form of the right side of Eq.(\ref{s11Lx}), but with subscripts $\mathcal{L}$ of $\langle S_{lr}\rangle$ and $\mathcal{H}$ in Eq.(\ref{s11Lx}) being replaced by subscripts $\mathcal{L'}$ and $\mathcal {H'}$.
Obviously, according to the definition of $\mathcal L$ and $\mathcal L'$, we immediately find that
\begin{equation}\label{eh} \mathcal H =\mathcal H'.\end{equation} We shall simply use the notation $\mathcal H$ for both of them.
Also, we can easily see that, constraints to $\langle S_{lr}\rangle_{\mathcal L'}$ listed in Eqs.(\ref{contr1z},\ref{controxz}) are identical to constraints to $\langle S_{lr}\rangle_{\mathcal{L}}$ as listed in Eqs.(\ref{contr1},\ref{controx}).
Therefore, given $\mathcal{H}$, the lower bound of $\underline{s}_{11}^{\mathcal L}$ calculated from Eq.(\ref{s11Lx}) and Eqs.(\ref{contr1},\ref{controx}) must be equal to the lower of $\underline{s}_{11}^{\mathcal{L'}}$ calculated from Eq.(\ref{s11Lz}) and Eqs.(\ref{contr1z},\ref{controxz}).
 And hence we can use only {\em one} functional form $\underline{\mathcal{S}}_{11}(\mathcal{H})$ for both the lower bound of $\underline{s}_{11}^{\mathcal{L}}$ and the lower bound of $\underline{s}_{11}^{\mathcal{L'}}$.

  Given theorem 2, we can actually deduce the lower bound of yield of  all single-photon pairs  basis through using the observed data in $X$ basis only, even for the non-asymptotic calculation. {\em This makes it possible to treat the yield of single-photon pairs and the phase-flip error of single-photon jointly} because both of them are dependent on the same quantity
  $\mathcal H$. This makes an important part to improve the efficiency of our key rates.



\begin{thebibliography}{99}
\bibitem{note1}There are other choices for set $\mathcal L$. For example, we can choose all pulses given by source $x,y,o$ i.e., $\mathcal{L} = \{oo,ox,xo,oy,yo,xx,yy,xy,yx\}$ and then use linear programming.
Here we take such a specific choice in order to show the main idea clearly with explicit formulas which can also make the calculation more efficiently.
Cautions should be taken that the computer can present wrong, insecure results quite often if linear programming is used.
\end{thebibliography}
\end{document}